\documentclass[aps,prl,twocolumn,showpacs,superscriptaddress,floatfix]{revtex4-1}
\usepackage{amsmath,amssymb}
\usepackage{graphicx}
\usepackage{epsfig}
\usepackage{color}
\usepackage{rotating}
\usepackage{bm}
\usepackage{setspace}

\begin{document}
\title{Electronic inhomogeneity and band structure on superstructural CuO$_2$ planes of infinite-layer Sr$_{0.94}$La$_{0.06}$CuO$_{2+y}$ films}
\author{Rui-Feng Wang}
\author{Jiaqi Guan}
\author{Yan-Ling Xiong}
\affiliation{State Key Laboratory of Low-Dimensional Quantum Physics, Department of Physics, Tsinghua University, Beijing 100084, China}
\author{Xue-Feng Zhang}
\affiliation{Institute of Physics, National Center for Electron Microscopy in Beijing, School of Materials Science and Engineering, Tsinghua University, Beijing 100084, China}
\author{Jia-Qi Fan}
\author{Jing Zhu}
\affiliation{Institute of Physics, National Center for Electron Microscopy in Beijing, School of Materials Science and Engineering, Tsinghua University, Beijing 100084, China}
\author{\\Can-Li Song}
\email[]{clsong07@mail.tsinghua.edu.cn}
\author{Xu-Cun Ma}
\email[]{xucunma@mail.tsinghua.edu.cn}
\affiliation{State Key Laboratory of Low-Dimensional Quantum Physics, Department of Physics, Tsinghua University, Beijing 100084, China}
\affiliation{Frontier Science Center for Quantum Information, Beijing 100084, China}
\author{Qi-Kun Xue}
\affiliation{State Key Laboratory of Low-Dimensional Quantum Physics, Department of Physics, Tsinghua University, Beijing 100084, China}
\affiliation{Frontier Science Center for Quantum Information, Beijing 100084, China}
\affiliation{Beijing Academy of Quantum Information Sciences, Beijing 100193, China}
\begin{abstract}
Scanning tunneling microscopy and spectroscopy are utilized to study the atomic-scale structure and electronic properties of infinite-layer Sr$_{0.94}$La$_{0.06}$CuO$_{2+y}$ films prepared on SrRuO$_3$-buffered SrTiO$_3$(001) substrate by ozone-assisted molecular beam epitaxy. Incommensurate structural supermodulation with a period of 24.5 $\textrm{\AA}$ is identified on the CuO$_2$-terminated surface, leading to characteristic stripes running along the 45$^\textrm{o}$ direction with respect to the Cu-O-Cu bonds. Spatially resolved tunneling spectra reveal substantial inhomogeneity on a nanometer length scale and emergence of in-gap states at sufficient doping. Despite the Fermi level shifting up to 0.7 eV, the charge-transfer energy gap of the CuO$_2$ planes remains fundamentally unchanged at different doping levels. The occurrence of the CuO$_2$ superstructure is constrained in the surface region and its formation is found to link with oxygen intake that serves as doping agent of holes in the epitaxial films.
\end{abstract}
\maketitle
\begin{spacing}{0.990}
High-temperature superconductivity in cuprates emerges upon doping an antiferromagnetic Mott insulator due to strong electron correlations \cite{Lee2006doping}. For understanding its mechanism and the emerging exotic phases (e.g. pseudogap and charge density waves) \cite{keimer2015quantum}, a central issue that must be clarified is how the ground state of the Mott insulator in the CuO$_2$ planes evolves with doping. In theory, it was often hypothesized that the doping induces significant spectral weight transfer from the high- to the low-energy scale \cite{Eskes1991anomalous, Meinders1993spectral, Senechal2000spectral} so that the ground state changes dramatically and some extraordinary electronic states develop near Fermi level ($E_\textrm{F}$) \cite{phillips2010colloquium}. This scenario has received some experimental supports from both bulk- and surface-sensitive measurements \cite{Uchida1991optical, Veenendaal1994electronic, Armitage2002doping, ye2013visualizing, cai2016visualizing}, and attracted increasing interest in the community of strongly correlated electron physics \cite{okada2013imaging}. However, the stability of the Zhang-Rice singlet with doping up to $x$ = 0.3 in La$_{2-x}$Sr$_x$CuO$_4$ poses a challenge to the prevailing view of spectral weight transfer \cite{Brookes2015stability}. The doping resulted changes in the electronic structure of the CuO$_2$ planes remain elusive in cuprates.

Structurally, all cuprates consist of alternating CuO$_2$ and various charge reservoir layers along the crystallographic $c$-axis \cite{park1995structures}. Superconductivity occurs in the CuO$_2$ planes when the chemical doping is implemented in the adjacent non-superconducting charge reservoir layers. In order to understand the physics of the superconducting CuO$_2$ planes and thus the pairing mechanism, it is highly tempting to investigate directly the CuO$_2$ planes in experiment \cite{misra2002atomic, Harter2012nodeless, Lv2015mapping, zhong2016nodeless}, provided the structural elegance and complexity of the cuprate superconductors. Indeed, direct measurement by scanning tunneling microscopy (STM) on the CuO$_2$ planes of infinite-layer Sr$_{1-x}$(La, Nd)$_{x}$CuO$_{2+y}$ films revealed a robust Mott-Hubbard band structure of CuO$_2$ against chemical doping \cite{Zhong2019direct}, which is in contrast to the usual assumption mentioned above. In this study, we investigate the hole-doped CuO$_2$ planes in a wide doping region by preparing epitaxial Sr$_{0.94}$La$_{0.06}$CuO$_{2+y}$ (SLCO) films on SrTiO$_3$ (STO) substrate with a SrRuO$_3$ (SRO) buffer layer, aiming to establish a comprehensive picture about the evolution of the ground state of the CuO$_2$ planes versus doping.

The experiments were conducted on a commercial ultrahigh vacuum (UHV) STM apparatus (Unisoku), which is connected to ozone-assisted molecular beam epitaxy (MBE) for \textit{in-situ} sample preparation. To reduce the lattice mismatch (1.2\%) between SLCO and STO, a buffer layer of 70-nm-thick SRO films that has a pseudo cubic lattice constant of 3.93 $\textrm{\AA}$, comparable to that (3.95 $\textrm{\AA}$) of SLCO, was firstly grown on Nb-doped STO(001) substrates using pulsed laser deposition technique. After transferred into UHV, the SRO-covered substrates were annealed at 500 $^\textrm{o}$C under ozone atmosphere to recover the atomically clean surface. The SLCO films were prepared by co-deposition of high-purity metals (Sr, La and Cu) from standard Knudsen cells under an ozone atmosphere of 1.1 $\times$ 10$^{-5}$ Torr, as detailed elsewhere \cite{Zhong2019direct}. Prior to STM measurements at 78 K, polycrystalline PtIr tips were calibrated on MBE-grown Ag/Si(111) epitaxial films. Tunneling spectra were measured using a standard lock-in technique with a small bias modulation at 931 Hz.
\end{spacing}

\begin{figure}[t]
\includegraphics[width=\columnwidth]{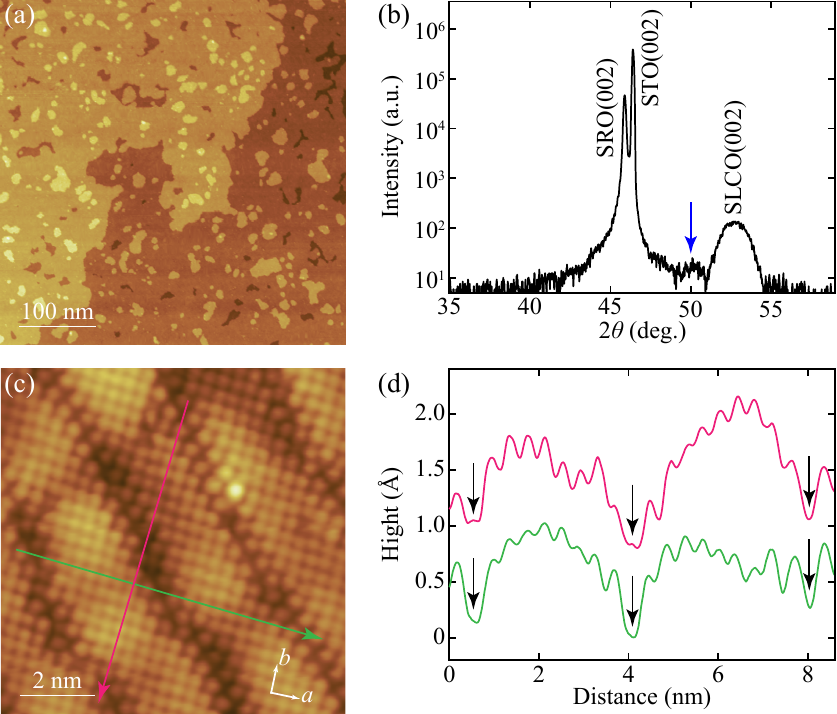}
\caption{(Color online) (a) STM topography of SLCO epitaxial film (450 nm $\times$ 450 nm, $V$ = -3.5 V, $I$ = 20 pA), decorated by small single-unit-cell islands. (b) XRD pattern around the SLCO(002) diffraction peaks measured using a monochromatic Cu K$_{\alpha1}$ radiation with a wavelength of 1.5406 $\textrm{\AA}$. (c) Atomic-resolved STM topographic image of superstructural CuO$_2$ (9.2 nm $\times$ 9.2 nm, $V$ = -0.85 V, $I$ = 30 pA). The bright spots correspond to the Cu atoms in the top layer. (d) Topographic profiles along the two Cu-O-Cu bond directions ($a$ and $b$), color coded to match the arrowed lines in (c). Black arrows mark the positions of the invisible Cu atoms.
}
\end{figure}

Figure 1(a) shows a large-scale STM topographic image of an as-prepared SLCO thin film with a thickness of 15 unit cells, in which the atomically-flat nature of the surface is apparent. The steps have a height of approximately 3.6 $\textrm{\AA}$, as expected for the infinite-layer SLCO and further supported by \textit{ex-situ} X-ray diffraction (XRD) measurement in Fig.\ 1(b). In addition to the (002) diffraction peaks of STO and SRO, electron-doped SLCO phase occurs predominantly with a $c$-axis lattice constant of 3.47 $\textrm{\AA}$. Meanwhile, a prominent new phase with a c-axis lattice constant of 3.6 $\textrm{\AA}$ (marked by blue arrow) appears and the phase turns out to be hole-doped SLCO near the sample surface, which will be discussed in detail below.

Illustrated in Fig.\ 1(c) is an atomically resolved STM topography of SLCO surface with an in-plane lattice parameter of 3.9 $\pm$ 0.1 $\textrm{\AA}$. Intriguingly, an incommensurate superstructure with a period of approximately 24.5 $\textrm{\AA}$ is observed, which runs along the diagonal direction of the CuO$_2$ square lattice and is very different from the primitive CuO$_2$(1 $\times$ 1) and reconstructed CuO$_2$(2 $\times$ 2) surfaces of the infinite-layer SLCO films on STO \cite{Zhong2019direct}. The superstructural CuO$_2$ planes are reminiscent of the well-known supermodulated BiO surfaces of Bi-family cuprates \cite{fischer2007scanning, misra2002atomic, Lv2015mapping, zhong2016nodeless}. The superstructure can be more clearly seen by line profiles along $a$ and $b$ axes in Fig.\ 1(d), where the black arrows denote the invisible atom rows along the [1\={1}0] direction. Such observation that the atoms in every eight or nine Cu atoms are invisible to STM is usually caused by structural displacement \cite{inoue1994study, massee2020atomic}, similar to the Bi-family cuprate superconductors \cite{shan2003stm, zou2020effect}. The structural supermodulation on the CuO$_2$ planes constitutes one of the main observations of this study.

\begin{figure*}[t]
\includegraphics[width=2\columnwidth]{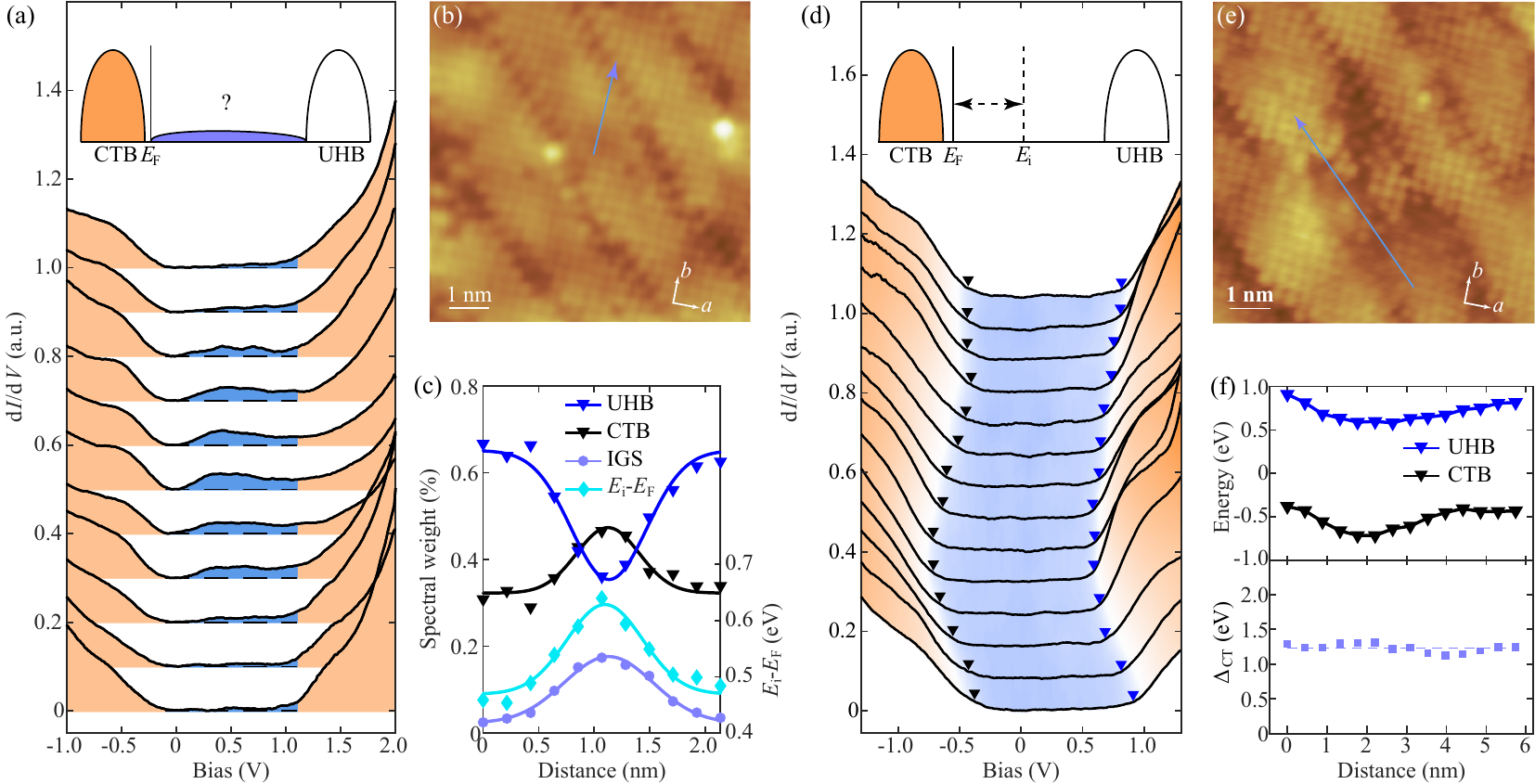}
\caption{(Color online) (a) Spatially resolved tunneling conductance \textit{dI/dV} spectra acquired along the blue arrow in (b). Color shaded areas measure spectral weights of CTB (left), IGS (middle) and UHB (right), respectively. Inset illustrates the schematic band structure of doped CuO$_2$. (b) STM topography of as-prepared SLCO (9.7 nm $\times$ 9.7 nm, $V$ = -1.2 V, $I$ = 10 pA). (c) Space-dependent variations in spectral weights and $E_\textrm{F}$ shift relative to $E_\textrm{i}$. (d) Spatially resolved \textit{dI/dV} spectra acquired along the blue arrow in (e). Black and blue triangles mark the onset energies of CTB and UHB, respectively. (e) STM topography (9.7 nm $\times$ 9.7 nm, $V$ = -0.8 V, $I$ = 10 pA) of UHV-annealed SLCO films at 510 $^\textrm{o}$C for one hour. (f) Space-dependent onset energies of UHB, CTB (top panel), and $\Delta_\textrm{CT}$ (bottom panel) in (d).
}
\end{figure*}

Next, we characterize the electronic structure of this newly observed superstructural CuO$_2$ surface by measuring a series of \textit{dI/dV} conductance spectra along a trajectory of 24.2 $\textrm{\AA}$ that almost covers the whole modulation period. The result is summarized in Fig.\ 2(a) and the corresponding surface is shown in Fig.\ 2(b). Evidently, all spectra are characterized by a charge-transfer gap (CTG) between two upturns in the density of the states. The two upturns correspond to the occupied charge transfer band (CTB) and the empty upper Hubbard band (UHB), respectively. The Fermi energy $E_\textrm{F}$ ($V$ = 0) is all close to the CTB, a hallmark of hole-doping \cite{Zhong2019direct}. Given that the substitution of trivalent La$^{3+}$ ions for Sr$^{2+}$ contributes to electron carriers, this unexpected finding implies that oxygen intake acts as a doping agent for holes in SLCO explored here. In contrast to the $p$-type SLCO films at $x >$ 0.1 \cite{Zhong2019direct}, the (2 $\times$ 2) superstructure caused by an appreciable intake and periodic occupation of apical oxygen atoms is absent in Fig.\ 2(b). This indicates a relatively lower oxygen doping, reconciling with our observation that $E_\textrm{F}$ is always located at an energy above CTB [Fig.\ 2(a)]. Furthermore, the spectra exhibit an obvious spatial inhomogeneity, as seen by the emergent in-gap states (IGS, blue shaded areas) within CTG. The IGS become more prominent in the bright regions in Fig.\ 2(b).

The spatial inhomogeneity becomes more evident when the spatial dependent spectral weights of CTB, IGS and UHB are deduced as the color shaded areas in Fig.\ 2(a). We show in Fig.\ 2(c) that the spectral weight of CTB, being in phase with that of IGS, increases with reducing UHB weight. Such a result seems understandable in the context of the scenario that spectral weight of CTB and IGS at lower energy builds up from a transfer from that of UHB at higher energy upon hole doping \cite{Eskes1991anomalous, Meinders1993spectral, Senechal2000spectral, phillips2010colloquium, Uchida1991optical, Veenendaal1994electronic, Armitage2002doping, ye2013visualizing, cai2016visualizing}. However, extreme caution should be taken, because the magnitude $\Delta_\textrm{CT}$ of CTG remains unchanged and there exists a systematic $E_\textrm{F}$ shift for different local doping \cite{Zhong2019direct}, albeit small in the heavily doped case (see the cyan diamonds in Fig.\ 2(c)). Given the fixed energy range during spectroscopic measurement, the hole doping induced $E_\textrm{F}$ downward shifting would naturally yield an inverse correlation between the space-dependent spectral weights of CTB and UHB in Fig.\ 2(c). In addition, heavier hole doping often means more IGS, thereby leading to a positive relationship between the local IGS and CTB.

To provide further insight into the origin of hole doping and IGS, we annealed the samples under UHV condition. Figure 2(d) represents the tunneling spectra along a trajectory of 52.1 $\textrm{\AA}$ on the surface of the annealed SLCO sample shown in Fig.\ 2(e). While the UHV annealing reduces the oxygen intake and shifts $E_\textrm{F}$ upwards \cite{Zhong2018continuous}, the electronic inhomogeneity becomes even more prominent: the brighter the STM contrast is, the smaller the energy separation between $E_\textrm{F}$ and CTB is. Similar to the previous report \cite{Zhong2019direct}, we determine the onset energies of CTB and UHB, as well as the separation $\Delta_\textrm{CT}$ between them [Fig.\ 2(f)] and find that CTB and UHB change in a synchronous manner so that $\Delta_\textrm{CT}$ remains essentially unchanged.

\begin{figure}[t]
\includegraphics[width=1\columnwidth]{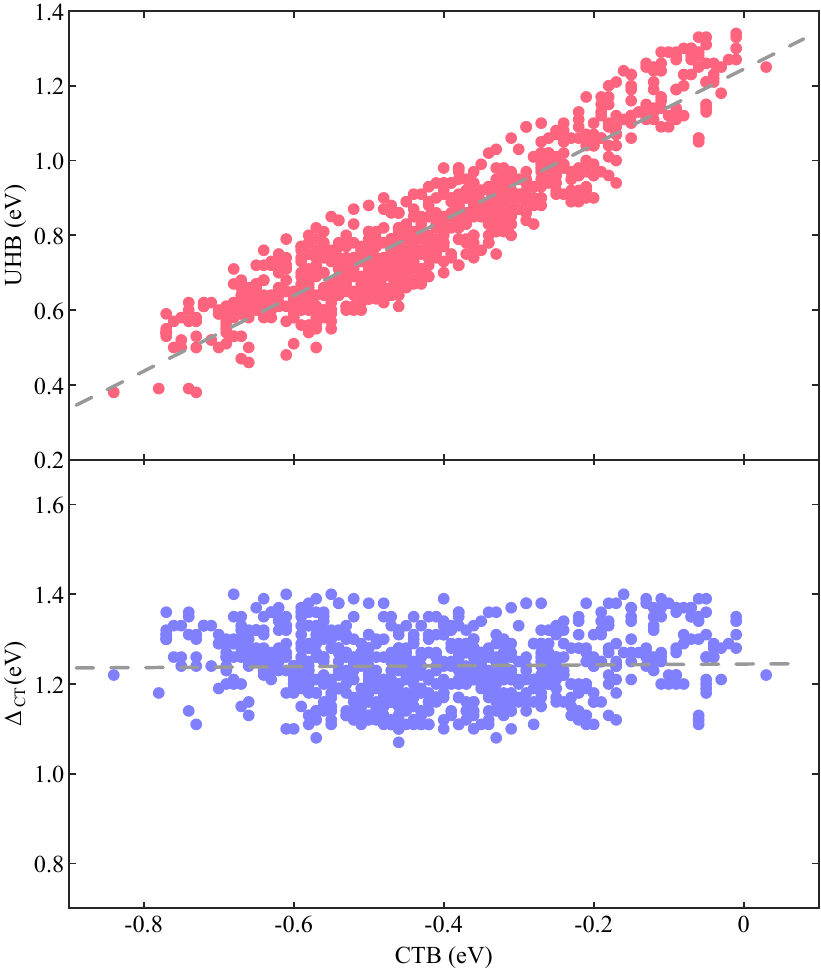}
\caption{(Color online) Onset energy of UHB (red circles) and $\Delta_\textrm{CT}$ (blue circles) plotted as a function of CTB onset energy. The statistics involve over 780 \textit{dI/dV} spectra at varied positions and samples. Gray dashed lines show the best linear fits to the data.
}
\end{figure}

\begin{figure}[t]
\includegraphics[width=\columnwidth]{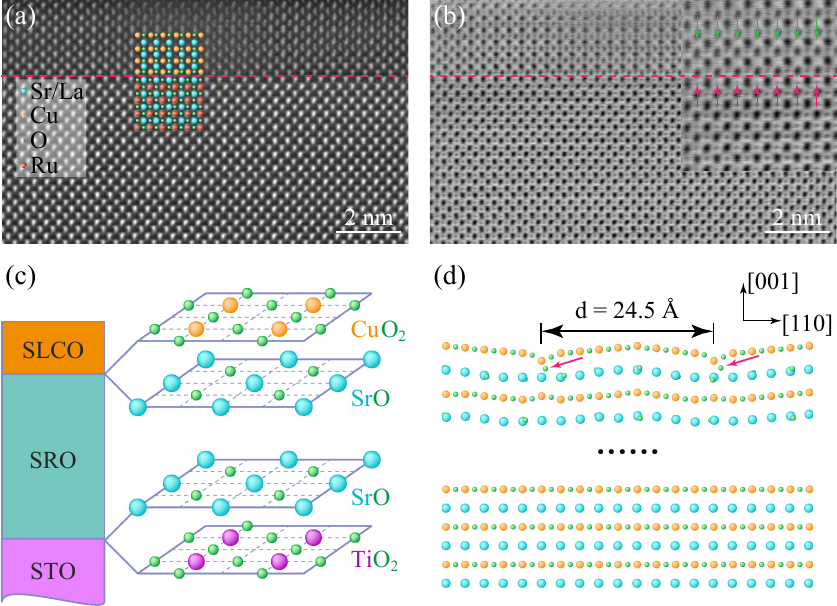}
\caption{(Color online) (a) HAADF-STEM image across the interface between SLCO and SRO along the [001] axis, marked by the dashed line. (b) ABF-STEM image in the same field of view as (a). Inset shows a zoom-in of the SLCO/SRO interface. The magenta arrows mark the oxygen in the SrO layer of SRO, while the blue arrows indicate no apical oxygen in the Sr/La layers of SLCO. (c) Schematic structure of superstructural SLCO films prepared on SRO-buffered STO substrates, with the interfacial stacking of SLCO/SRO and SRO/STO magnified. (d) Possible diagram of the superstructural SLCO near the surface. The magenta arrows mark the incorporated oxygens that serve as the doping agent of holes.
}
\end{figure}

We also note the significant suppression of IGS after UHV annealing, and ascribe it to the reduction of doping level. Thus, the underlying cause of spatial inhomogeneity and emergent IGS is obvious: it is the inhomogeneous distribution of local dopants (oxygen and La atoms), and this doping inhomogeneity gets more prominent after the UHV annealing. For increased doping of oxygen, $E_\textrm{F}$ gradually moves from the midgap energy $E_\textrm{i}$ (where the hole doping by oxygen compensates with the electron doping by La$^{3+}$)) to the CTB, whereas the fundamental Mott-Hubbard band structure remains intact. As the dopants are densely populated to a critical concentration, probably relating to the Bohr radius of the dopant atoms in question \cite{ALEXANDER1968semiconductor, Edwards1978university}, pronounced IGS or evanescent states emerge, prompting a transition from the Mott insulator to metallic or superconducting states. This bears a great similarity to the doping of semiconductors \cite{Van1992theory}. In any case, the fundamental Mott-Hubbard band structure of CuO$_2$ remains essentially unchanged, a hallmark of the self-modulation doping scheme \cite{Zhong2019direct}.

The robustness of Mott-Hubbard band structure against doping is further corroborated by annealing the SLCO films at different duration and measuring the corresponding conductance spectra in various regions. Figure 3 presents the extracted onset energies of UHB and $\Delta_\textrm{CT}$ as a function of the CTB onset energy. Compared to the CTB onset near $E_\textrm{F}$ on the as-grown SLCO film, the local CTB, or equivalently $E_\textrm{F}$, can shift continuously by 0.7 eV after UHV annealing. Surprisingly, the onset energy of UHB scales linearly with that of CTB, yielding a slope of 1.01, very close to unity. Consequently, the $\Delta_\textrm{CT}$ remains almost the same for all spectra we studied [bottom panel of Fig.\ 3]. The mean value of $\Delta_\textrm{CT}$ = 1.24 $\pm$ 0.07 eV turns out to be slightly smaller than that of SLCO films on STO substrates \cite{Zhong2019direct}. This might be caused by the reduced Madelung potential \cite{tsukada2006charge, ohta1991electronic}, probably owing to the slightly expanded in-plane lattice constant or the presence of structural supermodulation in the SLCO films on SRO. Nevertheless, the present study provides convincing experimental evidence that the doping changes little the fundamental Mott-Hubbard band structure of CuO$_2$, rather it only induces a systematic shifting of $E_\textrm{F}$ and IGS within CTG, as reported in $n$-type infinite-layer \cite{Zhong2019direct}.

Finally, we show by high-resolution scanning transmission electron microscopy (STEM) that the observed structural supermodulation occurs only in the surface regions of SLCO films. Figures 4(a) and 4(b) present the STEM images across the interface between the SLCO and SRO layer, taken in the high-angle annular dark-field (HAADF) and annular bright-field (ABF) modes, respectively. Evidently, the interface has a stacking sequence of RuO$_2$-SrO-CuO$_2$-Sr(La), as schematically drawn in Fig.\ 4(c). This is quite distinct from the SrO-TiO$_2$-Sr(La)-CuO$_2$ stacking for SLCO films grown directly on the STO substrates \cite{Zhong2019direct}. No structural supermodulation and apical oxygen atoms are observed in the bulk of the epitaxial SLCO films, in contrast to the Bi-family cuprates \cite{song2019visualization}. Actually, the bulk phase belongs to the well-established $n$-type SLCO \cite{Zhong2019direct} and contributes to the pronounced $n$-SLCO(002) diffraction peak seen in Fig.\ 1(b). Taken altogether, our results suggest that the superstructure should develop solely near the top surface region, as schematically illustrated in Fig.\ 4(d). The structural supermodulation forms to accommodate incorporation of oxygen atoms \cite{lv2016electronic}, which slightly expands the $c$-axis lattice constant, as we observe above.

In summary, our detailed STM investigation of a novel superstructured CuO$_2$ planes has provided information about the doping of cuprate superconductors. The unchanged Mott-Hubbard band structure and systematic shift of $E_\textrm{F}$, which is consistent with the self-modulation doping scheme, turn out to be the two primary features of the doping on CuO$_2$ planes, irrespective of the spatial electronic inhomogeneity and the varied doping levels. Such a simple scheme may be applicable to a number of other strongly correlated materials.

\begin{acknowledgments}
We acknowledge Pu Yu and Yingjie Lyu for providing the high-quality SRO/STO substrates. This work is financially supported by the Ministry of Science and Technology of China, the National Natural Science Foundation of China, and in part by the Beijing Advanced Innovation Center for Future Chip (ICFC).
\end{acknowledgments}

%

\end{document}